\documentclass[fleqn]{LxSymp_2026_Paper_Format}

\begin{document}

\title{A Physics-Informed, PIV-Based Method for Surface Pressure Reconstruction}
\firstpagehead{}
\runningheads{22nd LISBON Laser Symposium 2026}{}
\runnigfoots{}

\author{A. Daliri$^{\text{1},*}$, F. Scarano$^{\text{1}}$}

\address{1: Dept. of Aerospace Engineering, Delft University of Technology, The Netherlands\\
\vskip 1pt
{\addressize{*Corresponding author:
\href{mailto:a.daliri@tudelft.nl}{\textstyleInternetlink{a.daliri@tudelft.nl}}}}}
\fontfamily{phv}\fontseries{mc}\fontsize{9}{10}\selectfont
\linespread{1.2}
\vskip 3 mm
\begin{center}
\end{center}
\vskip 8 mm
\abstract{
A physics-informed method for PIV-based surface pressure reconstruction that combines least-squares optimization with both local and global physical constraints is proposed. The standard deviation of the surface--normal pressure gradient is introduced as a physically meaningful weighting function in the optimization process. Moreover, a physical global constraint is also introduced to reconstruct surface pressure outside the suction region. The results show very good agreement with data obtained from pressure taps. Near leading edge, the surface pressure reconstruction error has been reduced by one third. 
}

\section{Introduction}
\label{Sec:Intro}

Surface pressure measurement provides a direct link between flow physics and the resulting aero-hydrodynamic forces, moments, and energy transfer. Accurate surface pressure information is essential for validating numerical models, understanding flow separation and unsteady phenomena, and enabling reliable design and control of fluid--structure interaction systems. Common methods for surface pressure measurement include pressure-taps and pressure-sensitive paints (PSP) \citep{gregory2008review}. Pressure taps provide direct and accurate readings but are intrusive, limited in spatial resolution, and difficult to implement on complex 3D objects. PSP offers spatial measurements but requires careful calibration, is sensitive to temperature and lighting, and has limited temporal resolution. 

PIV-based pressure reconstruction techniques, which have been extensively worked on over the past decade, can overcome these limitations \citep{van2013piv}. They are non-intrusive, provides both spatially and temporally resolved pressure fields, and can capture unsteady flow phenomena, making it highly suitable for complex or dynamic aero-hydrodynamic studies. Thanks to the development of large-scale tomographic PIV \citep{scarano2012tomographic}, and 3D Lagrangian particle tracking \citep{schroder20233d}, the application of PIV-based pressure measurement techniques is increasingly moving from laboratory problems to real-world applications.

Accurate reconstruction of surface pressure from PIV data is a long-standing challenge in experimental fluid mechanics \citep{ragni2009surface,tagliabue2017surface,jux2020flow,cakir2024surface}. While PIV provides detailed velocity fields, pressure is not directly measurable and must be inferred through integration or optimization procedures that are highly sensitive to noise, spatial resolution, and boundary conditions. These difficulties become particularly severe in regions with strong pressure gradients, such as near stagnation points, suction peaks, and in the presence of flow separation or unsteady effects \citep{ragni2009surface,tagliabue2017surface}. As a result, conventional pressure reconstruction techniques often rely on ad-hoc smoothing and extrapolation , or other assumptions that limit their robustness and general applicability.

This work proposes a physics-informed method for PIV-based surface pressure reconstruction that combines least-squares optimization with both local and global physical constraints. The central idea is to exploit the statistical structure of the reconstructed pressure field normal to the surface. Specifically, the standard deviation of the pressure gradient along surface-normal lines is introduced as a physically meaningful weighting function in the optimization process, enabling the method to adaptively account for local uncertainty in the pressure reconstruction. Furthermore, the same statistical indicator is used as an objective criterion to identify key aerodynamic features, including stagnation points and suction peaks, directly from PIV data. This unified framework improves robustness, reduces user intervention, and enhances the physical consistency of surface pressure reconstruction from experimental velocity measurements.

In the following sections, first the proposed PIV-based surface pressure reconstruction method is described. Then, results for the case of an airfoil is presented and discussed. Finally, a summary of key findings and insights are given.

\section{Surface Pressure Reconstruction Method}
\label{Sec:method}

Although direct evaluation of surface pressure using simple linear \citep{ragni2009surface} or second-order \citep{tagliabue2017surface} extrapolation estimates surface pressure with good accuracy in a large portion of the surface, it has shown that even in cases where the boundary layer thickness is very small (e.g. transonic flow), it is associated with considerable error in regions with high pressure gradients such as suction peaks and stagnation points, where uncertainties due to measurement noise and spatial resolution are highest. As a matter of fact, these regions both contain the most interesting physics phenomena and contribute significantly to the aerodynamic forces. Therefore, one must be able to estimate surface pressure in these regions with high accuracy. To address this challenge, the present method formulates surface pressure reconstruction as an optimization problem, in which physically informed constraints are used to identify the most plausible pressure distribution consistent with the measured flow--field. 

\begin{figure}[htbp]
	\centering
	\includegraphics[width=0.85\textwidth]{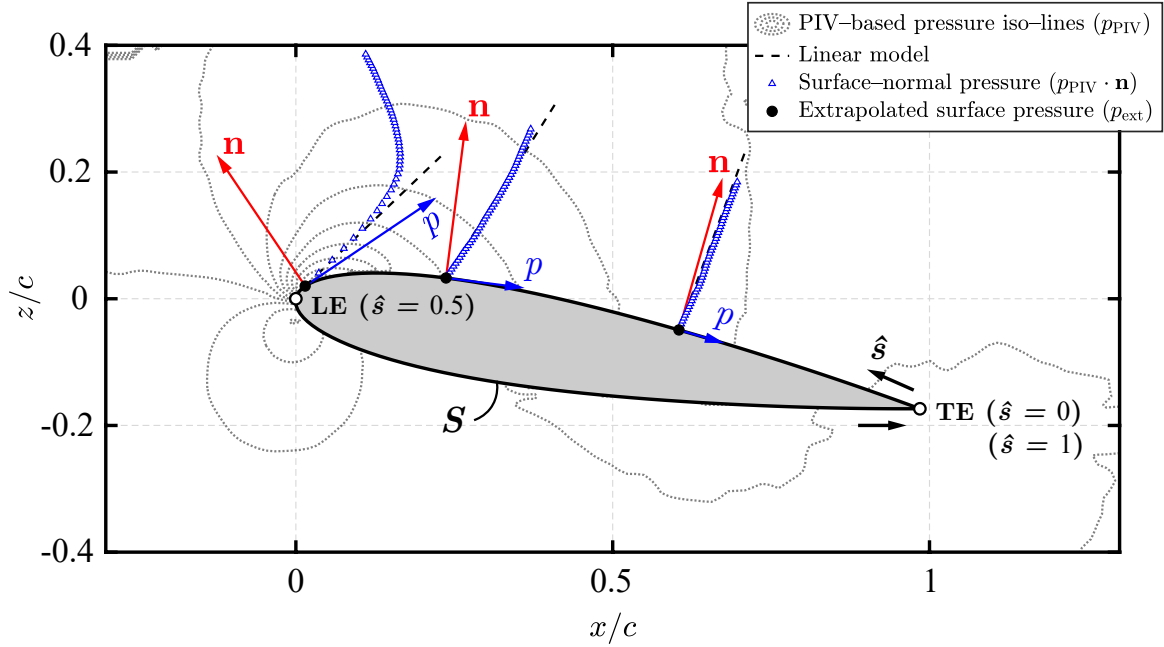}
	\centering
	\caption{Schematic of airfoil model along with illustration of parameters of surface pressure reconstruction method}
	\label{fig:schematic}
\end{figure}

The details of this optimization problem are described in the following. Figure. \ref{fig:schematic} shows a schematic of a NACA 0015 airfoil at an angle of attack of 10 degrees. This schematic, which is consistent with the dataset that will be examined in this paper, shows the parameters related to the surface pressure reconstruction method that will be explained.

\subsection{Cost Function Formulation}
\label{Sec:cost}

Let $p_S(\tilde{s})$ denote the surface pressure distribution along the airfoil surface ($S$) parameterized by the normalized curvilinear coordinate
$\tilde{s}$ (Fig. \ref{fig:schematic}). $\tilde{s}$ originates from trailing edge (TE) and for a symmetric surface (e.g. NACA0015), at the leading edge (LE), $\tilde{s}=0.5$. The optimization is performed directly on the surface pressure ($p_S$), which improves numerical stability and facilitates the imposition of surface-based physical constraints. The cost function is defined as,

\begin{ceqn}
	\begin{align}
		J = \sum_{i=1}^{N_s} w_i \left[ p_S(\tilde{s}_i) - p_{\mathrm{ext}}(\tilde{s}_i) \right]^2
		+ \lambda \int_{S} \left( \frac{\partial^2 p_S}{\partial \tilde{s}^2} \right)^2 \, d\tilde{s} .
		\label{Eq:cost}
	\end{align}
\end{ceqn}

where,
\begin{itemize}
	\item $p_S(\tilde{s})$ is the surface pressure distribution represented by B-spline basis functions \citep{piegl1995b};
	\begin{ceqn}
		\begin{align}
			p_s(\tilde{s}) = \sum_{k=1}^{M} c_k \, B_k(\tilde{s}),
			\label{Eq:bspline}
		\end{align}
	\end{ceqn}
	where $B_k(\tilde{s})$ are B-spline basis functions, $c_k$ are the corresponding control coefficients and $M$ is the number of basis functions. This representation ensures inherent smoothness of the solution and allows the derivatives to be evaluated analytically and efficiently. During each iteration of the optimization process, the B-spline coefficients $c_k$ are updated to minimize the cost function, resulting in a robust and computationally efficient reconstruction of surface pressure. $M$ should be chosen such that the B-spline can reconstruct the peaks and at the same time does not lead to large fluctuations in the reconstructed pressure.
	 
    \item $p_{\mathrm{ext}}(\tilde{s})$ is the linearly extrapolated surface pressure at location $\tilde{s}_i$. To calculate this, The pressure field obtained from PIV ($p_\mathrm{PIV}$) is first reconstructed in the near-wall flow region using the Navier--Stokes equations.
    \begin{ceqn}
    	\begin{align}
    		\nabla{p} = -\rho\frac{D\mathbf{u}}{Dt}+\mu\nabla^2\mathbf{u} .
    		\label{Eq:NS}
    	\end{align}
    \end{ceqn}
    where $\rho$ and $\mu$ are the mass density and the dynamic viscosity, respectively. For details on how to calculate pressure from PIV, refer to \citet{van2013piv}. In current work, for calculating pressure form PIV, iterative domain partitioning (IDP) method \citep{Dontu2026idp} is used to minimize the integration error propagation and dependence on boundary condition. Figure \ref{fig:schematic} shows an illustration of PIV--based pressure iso--lines for the airfoil at incidence. 
    
    Then, at each surface location $\tilde{s}_i$, pressure values along the corresponding surface--normal direction $\mathbf{n}$ ($p_\mathrm{PIV}\cdot\mathbf{n}$) are linearly extrapolated toward the wall to estimate the extrapolated surface pressure $p_{\mathrm{ext}}(\tilde{s}_i)$ (see, Fig. \ref{fig:schematic}). Because this extrapolated pressure is sensitive to experimental uncertainties and local flow complexity, it is not imposed directly as the final solution, but rather incorporated into the cost function in a weighted least-squares sense. This procedure provides a physically grounded but potentially inaccurate estimate of surface pressure, which serves as the reference in the optimization process.
    
    \item $w_i$ is a spatially varying weighting function. This weighting function controls the relative influence of the extrapolated surface pressure, $p_{\mathrm{ext}}(\tilde{s}_i)$, at each surface location in the optimization, allowing regions with higher confidence to contribute more strongly to the reconstructed surface pressure. Larger weights enforce closer agreement with the extrapolated surface pressure, while smaller weights relax this constraint and allow the regularization term to dominate, improving robustness in regions of higher uncertainty or complex flow behavior. This function is calculated using the surface-normal pressure gradient ($\nabla{p_\mathrm{PIV}}\cdot\mathbf{n}$), which will be explained later. It will be shown later that this parameter will also be used to identify the suction peak location.
    \item To ensure physical consistency and suppress non--physical oscillations, a regularization term based on the second--derivative of surface pressure is introduced. Penalizing $\partial^2{p_S}/\partial{\tilde{s}^2}$, discourages unrealistically sharp surface pressure fluctuations while preserving genuine aerodynamic features such as suction peaks. 
    \item $\lambda$ is a regularization parameter, which controls the balance between fidelity to the extrapolated pressure data and smoothness of the reconstructed surface pressure. It is used to stabilize the solution and prevent non--physical oscillations caused by noise or ill-posedness.
\end{itemize}

In general, the first term enforces consistency between the optimized surface pressure and the linearly extrapolated surface pressure, while the second term acts as a surface smoothness constraint, reflecting the physical expectation that surface pressure varies smoothly along the surface except in regions of strong flow features. For these special cases, the regularization parameter $\lambda$ can be defined as a function of the surface location and the values can be modified in the estimated regions of discontinuity (like shock wave region) to prevent the elimination of physical phenomena. For now, in current work, $\lambda$ is considered to be a constant.

\subsection{Surface--normal pressure gradient}
\label{Sec:std}

As can be seen from Fig. \ref{fig:schematic}, the variations of surface--normal pressure near the wall are very severe in the suction peak and stagnation point regions. Moving from the leading edge towards the trailing edge, these changes become less significant. This is the reason for the large error of linear extrapolation in these regions. Linear model cannot follow such sharp gradients. It has been shown that although higher order models give better results in these areas, still have significant errors because the rate of pressure changes near the surface increases as one approaches the wall. One way to quantify this effect is to use the surface--normal pressure gradient ($\nabla{p_\mathrm{PIV}}\cdot\mathbf{n}$). In this method, the standard deviation of the surface--normal pressure gradient ($\sigma_n$) is used as a statistical variable to evaluate the pressure gradient changes and identify high gradient regions;
 \begin{ceqn}
	\begin{align}
		\sigma_n(\tilde{s}_i) = 
		\sqrt{\frac{1}{N}\int_{0}^{N}\left(\nabla{p_\mathrm{PIV}\cdot\mathbf{n}(\tilde{s}_i)}-\overline{\nabla{p_\mathrm{PIV}\cdot\mathbf{n}(\tilde{s}_i)}}\right)\cdot{dn}}.
		\label{Eq:sigma}
	\end{align}
\end{ceqn}
where $n$ is the distance from the surface along $\mathbf{n}$ , $N$ is the chosen normal--line length and $\overline{\nabla{p_\mathrm{PIV}\cdot\mathbf{n}(\tilde{s}_i)}}$ is the mean gradient along $\mathbf{n}$. This quantity serves as a physically meaningful indicator of local uncertainty: regions with higher $\sigma_n$ correspond to larger local variations in the pressure field, such as near stagnation points, suction peaks, or separation zones.

In the optimization, $\sigma_n$ is used to calculate the weighting function $w_i$ in Eq. \ref{Eq:cost};
  \begin{ceqn}
 	\begin{align}
 		w_i = 
 		\frac{1}{\sigma_n(\tilde{s}_i)+\epsilon}.
 		\label{Eq:weight}
 	\end{align}
 \end{ceqn}
where, $\epsilon$ is a small constant to prevent division by zero. This strategy emphasizes regions with more stable, physically consistent gradients, while allowing the method to remain robust in high gradient or noisy regions.

In addition to weighting the optimization, $\sigma_n$ can be used as an objective criterion to identify key aerodynamic features directly from the PIV data. For instance, peaks in 
$\sigma_n$ often correspond to local extrema in surface pressure, enabling the detection of stagnation points, suction peaks, and regions of flow separation without manual intervention. This unified framework reduces user bias, enhances physical consistency, and improves the robustness of surface pressure reconstruction from experimental velocity measurements. It will be shown that $\sigma_n$ distribution on the airfoil surface is used to identify the suction peak and can identify suction peak location with a  reasonable accuracy.

\subsection{Optimization Constraints}
\label{Sec:constraints}

The reconstruction of surface pressure is governed by a set of local and global physical constraints that ensure consistency with fundamental aerodynamic principles. Local constraints enforce physically admissible behavior near critical regions, such as the leading edge, where the pressure gradient must remain bounded, and at stagnation points, where by definition the pressure coefficient is equal to unity. In addition, a global constraint is imposed providing an overall balance condition that anchors the reconstructed pressure field and enforces coherence between local surface behavior and the global flow physics.

In current work, a global constraint is proposed that is directly derived from the Navier-Stokes equation. It can be shown that considering a control volume $CV$ in the flow--field that encompasses the surface of the object $S$, the following relation is true. For sake of brevity, proof is not given.
  \begin{ceqn}
	\begin{align}
		\iint_S\nabla{p_S}\cdot\mathbf{n}~d\tilde{s} =
		-\iint_S\left[\nabla{\left(\frac{1}{2}\rho|\mathbf{u}_S|^2\right)+\rho(\omega_S\times\mathbf{n})}\right]\cdot\mathbf{n}~ds\notag\\
		-\iint_{CS}\left[\nabla(p+q)+\rho(\omega\times\mathbf{n})\right]\cdot\mathbf{n}~ds
		\label{Eq:constraint}
	\end{align}
\end{ceqn}
where, $CS$ is the surface of control volume and $\mathbf{u}_S$ and $\omega_S$ are translational and rotational velocity vectors of the object's surface, respectively. $p$ and $q$ are static and dynamic pressure of flow, respectively. If the object surface is fixed or the object is solid and the origin of coordinate system is on the object surface, the first term on the right-hand side will be equal to zero.

\section{Results and Discussions}
\label{Sec:results}

The proposed surface pressure reconstruction method is applied to time-averaged PIV dataset obtained for a NACA 0015 airfoil at an angle of attack of $10^\circ$ and a Reynolds number of 200,000. The measurements were acquired using three-dimensional Lagrangian Particle Tracking (3D-LPT) in the open-jet wind tunnel facility at Delft University of Technology, providing volumetric velocity fields. The airfoil model was bounded from both ends and the flow in the measurement volume was two-dimensional (2D). So, the velocity data is averaged along the span to obtain a high quality 2D flow--field. The pressure field is reconstructed by integrating Eq. \ref{Eq:NS} and using iterative-domain partitioning (IDP) technique. Further details regarding the experimental setup, measurement procedure, and data processing can be found in \citep{Dontu2026idp}. 

\begin{figure}[htbp]
	\centering
	\includegraphics[width=1\textwidth]{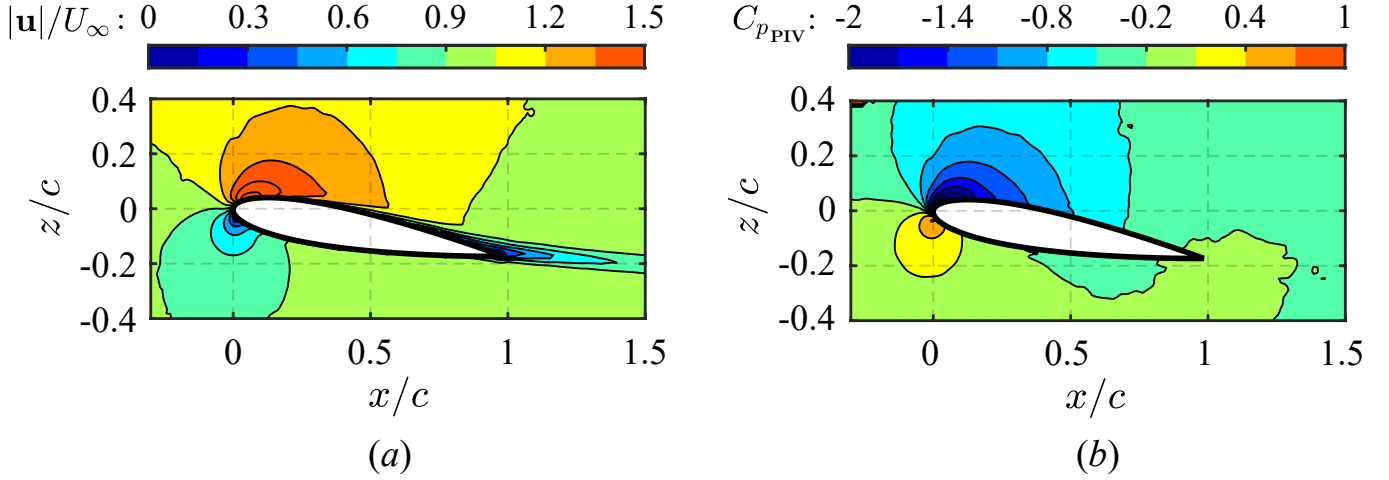}
	\centering
	\caption{a) non--dimensional velocity magnitude from PIV,~~b) pressure coefficient field reconstructed from PIV}
	\label{fig:u-cp-contour}
\end{figure}

Figure \ref{fig:u-cp-contour} shows a representative time- and span--averaged non--dimensional velocity magnitude field together with the corresponding pressure coefficient contours reconstructed from the PIV data, serving as the basis for the surface pressure reconstruction analysis discussed in the following sections.

\subsection{Surface--normal pressure gradient}
\label{Sec:pgradresult}

\begin{figure}[htbp]
	\centering
	\includegraphics[width=1\textwidth]{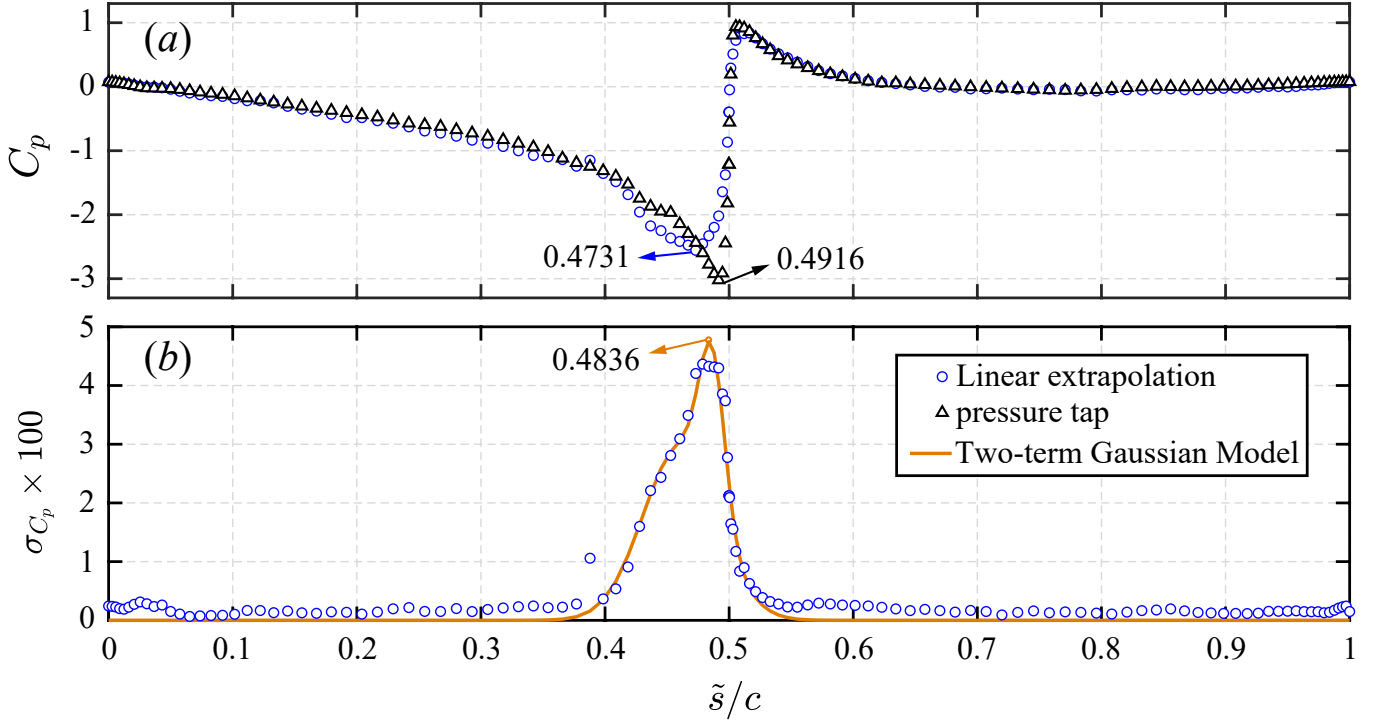}
	\centering
	\caption{a) surface pressure coefficient,~~b) standard deviation of the surface--normal pressure coefficient gradient}
	\label{fig:cp-sigma}
\end{figure}

Figure \ref{fig:cp-sigma}-a Compares the pressure coefficients obtained from linear interpolation with the pressure coefficients obtained from pressure taps. It clearly shows that near the leading edge, the pressure coefficient obtained from the linear extrapolation has a considerable error. The location of the suction peak and the value of the maximum suction are not estimated correctly. Figure \ref{fig:cp-sigma}-b shows the standard deviation distribution of surface--normal pressure gradient. It confirms that the standard deviation is significantly high near the leading edge. Therefore, the standard deviation can be a good measure for identifying areas of high uncertainty.

On the other hand, Figure 3 shows that the maximum standard deviation occurs almost at the suction peak location. A two-term Gaussian function is fitted to the standard deviation distribution ​​to obtain the accurate location of the maxima. The position of the maximum of the Gaussian function is very close to the suction peak location obtained from the pressure taps. Gaussian function identifies this location with an error of 1.5\%. The suction peak location obtained from the linear extrapolation has an error of 3.5\%. The governed dataset is of high quality and there is no significant reflections on the airfoil surface. So it should be noted that in case of reflection from the surface, this error will be much higher for linear extrapolation. Thus, the results confirm that the standard deviation is a suitable tool for identifying the suction peak location as well.

\subsection{Global Constraint Evaluation}
\label{Sec:constraineval}

\begin{figure}[htbp]
	\centering
	\includegraphics[width=1\textwidth]{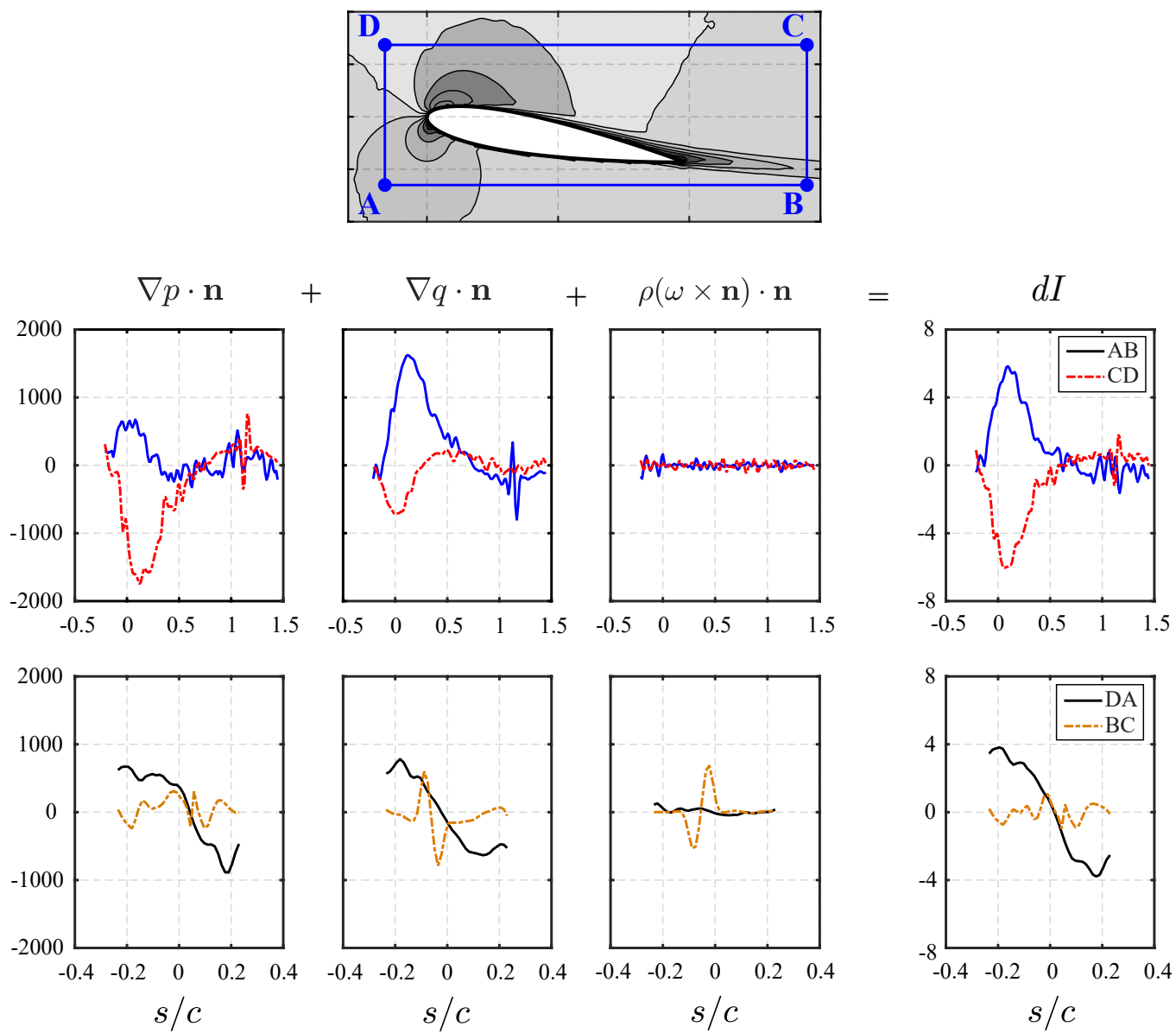}
	\centering
	\caption{Comparison of different terms of global constraints on the control--volume surface}
	\label{fig:cv-terms}
\end{figure}

Figure \ref{fig:cv-terms} compares different terms of Eq. \ref{Eq:constraint} calculated using the velocity field obtained from PIV and the corresponding reconstructed pressure field on a control--volume ABCD. As the figure shows and can be mathematically proven for the general case, the distribution of the different terms of Eq. \ref{Eq:constraint} on the boundaries of the control--volume is such that the total integral will ultimately be equal to zero. Despite the errors in velocity measurement and pressure calculation, the exact value of the right-hand side integral of Eq. \ref{Eq:constraint} for the present case is equal to 0.036, which is approximately zero. So, the simpler form of the global constraint is;

  \begin{ceqn}
	\begin{align}
		\iint_S\nabla{p_S}\cdot\mathbf{n}~d\tilde{s} =
		0
		\label{Eq:constraint0}
	\end{align}
\end{ceqn}

\subsection{Reconstructed Surface pressure}
\label{Sec:Cprec}

\begin{figure}[htbp]
	\centering
	\includegraphics[width=1\textwidth]{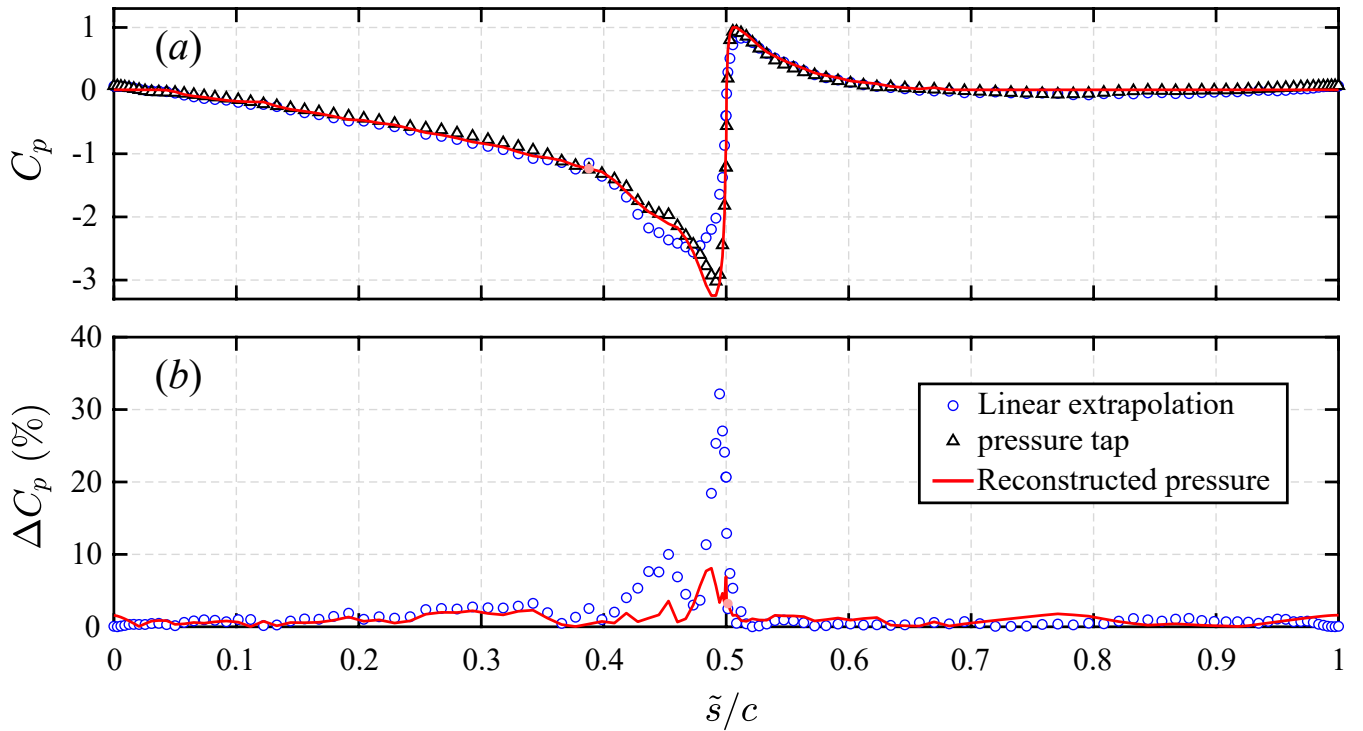}
	\centering
	\caption{a) surface pressure coefficient,~~b) relative error of surface pressure coefficient}
	\label{fig:cp-error}
\end{figure}

Figure \ref{fig:cp-error} shows the surface pressure reconstruction result. This figure shows that the applied method has been able to recover the suction peak effectively and reconstruct the location and magnitude of the suction peak with reasonable accuracy. Interestingly, the method can even reconstruct other flow features, including the flow transition effects, which is very hard to determine with conventional methods. The accuracy of pressure calculation using the proposed method in the areas adjacent to the leading edge has been improved by more than 3 times, which is a promising result.

\begin{figure}[htbp]
	\centering
	\includegraphics[width=1\textwidth]{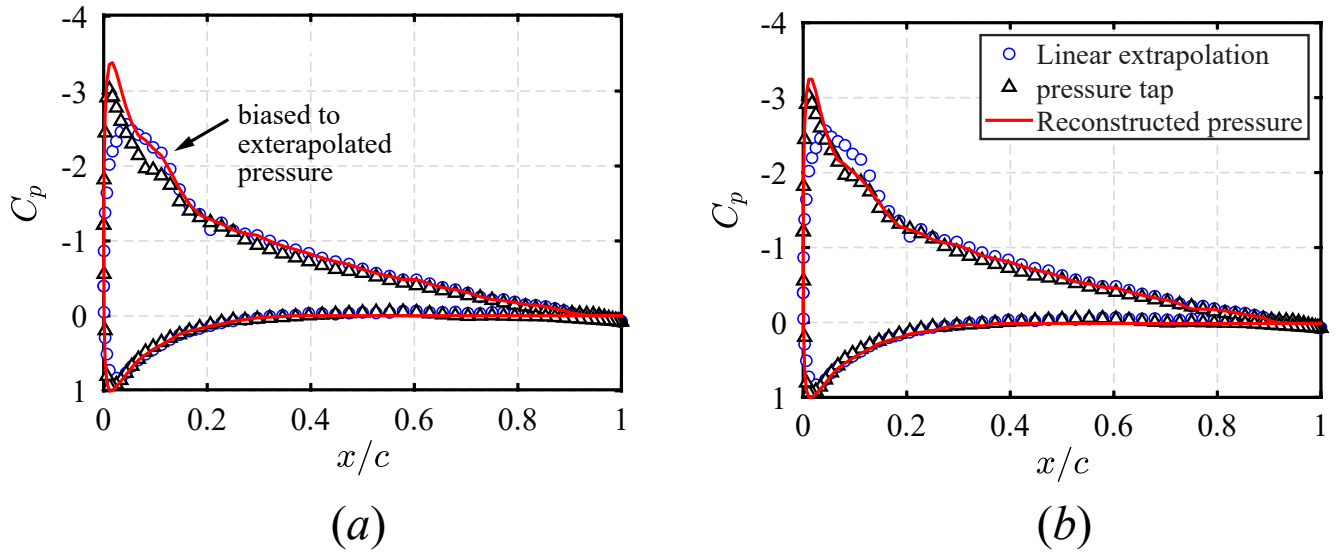}
	\centering
	\caption{a) reconstructed surface pressure coefficient without global constraint ,~~b) reconstructed surface pressure coefficient with global constraint}
	\label{fig:cp-xc}
\end{figure}

Figure \ref{fig:cp-xc} compares surface pressure distribution on the airfoil chord with and without employing global constraint. Although in both cases the other constraints mentioned in sections \ref{Sec:constraints} and \ref{Sec:pgradresult} are successful in recovering the pressure in the suction region, in the case where the global constraint is not used, the data outside the suction peak region shows a strong bias towards the pressure obtained from the linear extrapolation. It can be concluded that the use of the global constraint of Eq. \ref{Eq:constraint0} is necessary to reconstruct phenomena outside the suction peak region that may not be accounted for by the linear extrapolation for some reasons.

\section{Conclusions}
\label{Sec:Conclusions}

A physics-informed method for PIV-based surface pressure reconstruction that combines least-squares optimization with both local and global physical constraints is proposed. The standard deviation of the surface--normal pressure gradient is introduced as a physically meaningful weighting function in the optimization process, enabling the method to adaptively account for local uncertainty in the pressure reconstruction. Furthermore, the same statistical indicator is used as an objective criterion to identify key aerodynamic features, including suction peak location, directly from PIV data. Also,  a global constraint is proposed that is directly derived from the Navier-Stokes equation that enforces coherence between local surface behavior and the global flow physics. The results show very good agreement between the reconstructed surface pressure with one obtained from pressure taps. Near leading edge, the surface pressure reconstruction error has been reduced by one third.

\bibliography{LxSymp2026_Bib}
\bibliographystyle{apacite}


\end{document}